\documentclass[a4paper,11pt]{article}
\usepackage{pos}

\newcommand{\bea}{\begin{eqnarray}}
\newcommand{\eea}{\end{eqnarray}}
\newcommand{\be}{\begin{equation}}
\newcommand{\ee}{\end{equation}}

\newcommand{\ar}{a_s}

\title{Analytic QCD: Recent Results}

\author*[a]{A.V.~Kotikov}
\author[b,c]{I.A.~Zemlyakov}

\affiliation[a]{Joint Institute for Nuclear Research,\\ 141980, Dubna, Moscow region, Russia}

\affiliation[b]{Department of Physics, Universidad Tecnica Federico Santa Maria,\\ Avenida Espana, Valparaiso 1680, Chile}

\affiliation[c]{Tomsk State University, 634010 Tomsk, Russia}

\emailAdd{kotikov@theor.jinr.ru}
\emailAdd{zemlyakov@theor.jinr.ru}

\abstract{
  We present a brief overview of analytical QCD, focusing primarily on a less common form of the analytical coupling $A_{\rm MA}(Q^2)$, which is particularly convenient for $Q^2\sim\Lambda^2$.
  This form has been extensively used in recent studies of the (polarized) Bjorken sum rule and the Gross-Llewellyn Smith sum rule.}

\FullConference{QCD at the Extremes (QCDEX2025)\\
1–5 Sept 2025\\
Online\\}

\begin{document}
\maketitle

\label{sec:intro}
\section{Introduction}
The strong coupling constant $\alpha_s(Q^2)$ satisfies the renormalization group equation
\be
L\equiv \ln\frac{Q^2}{\Lambda^2} = \int^{\overline{a}_s(Q^2)} \, \frac{da}{\beta(a)},~~ \overline{a}_s(Q^2)=\frac{\alpha_s(Q^2)}{4\pi}\,,~~\ar(Q^2)=
\beta_0\,\overline{a}_s(Q^2)\,,
\label{RenGro}
\ee 
with a specific boundary condition and the QCD $\beta$-function:
\be
\beta(\ar)~=~
-\beta_0  \overline{a}_s^{2} \, \Bigl(1+\sum_{i=1} b_i \ar^i \Bigr),~~
b_i=\frac{\beta_i}{\beta_0^{i+1}}\,, ~~
 \beta_0=11-\frac{2f}{3},~~\beta_1=102-\frac{38f}{3},
\label{beta}
\ee
for $f$ active quark flavors. Currently, the first five coefficients $\beta_i$ ($i\leq 4$) are known exactly \cite{Baikov:2016tgj}.
In the present analysis, we will only require $i=0$ and $i=1$.

For $Q^2 \gg \Lambda^2$, Eq. (\ref{RenGro}) can be solved iteratively in the form of a $1/L$-expansion, which can be compactly expressed as:
\be
a^{(1)}_{s,0}(Q^2) = \frac{1}{L_0},~~
a^{(2)}_{s,1}(Q^2) = 
a^{(1)}_{s,1}(Q^2) +
\delta^{(2)}_{s,1}(Q^2)
\label{as}
\ee
where the next-to-leading order (NLO) correction is:
\be
\delta^{(2)}_{s,k}(Q^2) = - \frac{b_1\ln L_k}{L_k^2} ,~~
L_k=\ln t_k,~~t_k=\frac{1}{z_k}=\frac{Q^2}{\Lambda_k^2}\,.
\label{L}
\ee

Thus, already at leading order (LO), where $\ar(Q^2)=\ar^{(1)}(Q^2)$, and in any order of perturbation theory (PT), the coupling $\ar(Q^2)$ incorporates its own dimensional transmutation parameter $\Lambda$, which is related to the normalization $\alpha_s(M_Z^2)$. The value $\alpha_s(M_Z)=0.1180$ is given in PDG24 \cite{PDG24} (see also \cite{Enterria}).

{\bf $f$-dependence of the coupling $\ar(Q^2)$.}~~
The coefficients $\beta_i$ in (\ref{beta}) depend on the number $f$ of active quarks, which affects the coupling $\ar(Q^2)$ at threshold values $Q^2_f \sim m^2_f$, where additional quarks become active for $Q^2 > Q^2_f$.
Consequently, the coupling $a_s$ depends on $f$, and this dependence is incorporated into $\Lambda$, i.e., $\Lambda^f$ appears in Eqs. (\ref{RenGro}) and (\ref{as}).

The relationship between $\Lambda_{i}^{f}$ and $\Lambda_{i}^{f-1}$ is known up to the four-loop order \cite{Chetyrkin:2005ia} in the $\overline{MS}$ scheme.
Here, we will not address the $f$-dependence of $\Lambda_{i}^{f}$, as we are primarily interested in the low-$Q^2$ region and therefore use $\Lambda_{i}^{f=3}$ (see, e.g., \cite{Chen:2021tjz}):
\be
\Lambda_0^{f=3}=142~~ \mbox{MeV},~~\Lambda_1^{f=3}=367~~ \mbox{MeV}
\,.
\label{Lambdas}
\ee

\section{Fractional Derivatives}

Following \cite{Cvetic:2006mk,Cvetic:2006gc}, we define the derivatives (at the $(i)$-th order of PT) as:
\be
\tilde{a}^{(i)}_{n+1}(Q^2)=\frac{(-1)^n}{n!} \, \frac{d^n a^{(i)}_s(Q^2)}{(dL)^n} \, ,
\label{tan+1}
\ee
which are particularly useful in the context of analytic QCD (see, e.g., \cite{Kotikov:2022JETP}).

The series of derivatives $\tilde{a}_{n}(Q^2)$ can effectively replace the corresponding series of powers of $\ar$. Each derivative reduces the power of $\ar$ but introduces an additional factor of the $\beta$-function $\sim \ar^2$. Thus, each derivative effectively adds a factor of $\ar$, making it feasible to use derivative series in place of power series.

At LO, the derivative series $\tilde{a}_{n}(Q^2)$ coincide exactly with $\ar^{n}$. Beyond LO, the relationship between $\tilde{a}_{n}(Q^2)$ and $\ar^{n}$ was established in \cite{Cvetic:2006gc,Cvetic:2010di} and extended to non-integer values $n \to \nu$ in \cite{GCAK}.

Now consider the $1/L$-expansion of $\tilde{a}^{(k)}_{\nu}(Q^2)$ $(k=0,1)$ at LO and NLO:
\be
\tilde{a}^{(1)}_{\nu,0}(Q^2)={\bigl(a^{(1)}_{s,0}(Q^2)\bigr)}^{\nu} = \frac{1}{L_0^{\nu}},~
\tilde{a}^{(2)}_{\nu,1}(Q^2)=\tilde{a}^{(1)}_{\nu,1}(Q^2) +
\nu\, \tilde{\delta}^{(2)}_{\nu,1}(Q^2),~~
\label{tdmp1N}
\ee
where
\be
\tilde{\delta}^{(2)}_{\nu,1}(Q^2)=
\hat{R}_1 \, \frac{1}{L_i^{\nu+1}}= \Bigl[\hat{Z}_1(\nu)+ \ln L_i\Bigr]\, \frac{1}{L_i^{\nu+1}},~~
\hat{R}_1=b_1 \Bigl[\hat{Z}_1(\nu)+ \frac{d}{d\nu}\Bigr],~~\hat{Z}_1(\nu)=\Psi(\nu+1)+\gamma_{E}-1
\label{hR_i}
\ee
with $\gamma_{E}$ being Euler's constant and $\Psi(\nu+1)$ the $\Psi$-function.

The representation (\ref{tdmp1N}) of the $\tilde{\delta}^{(2)}_{\nu,1}(Q^2)$ correction in terms of the $\hat{R}_1$ operator\footnote{Operators similar to $\hat{R}_1$ were previously used in \cite{BMS1}.} is crucial and allows for a similar representation of higher-order results in the $1/L$-expansion of analytic couplings.

\section{MA Coupling}

In \cite{ShS}, an effective approach was developed to eliminate the Landau singularity, based on a dispersion relation that connects the new analytic coupling $A_{\rm MA}(Q^2)$ with the spectral function $r_{\rm pt}(s)$ derived from PT. At LO, this gives:
    \be
A^{(1)}_{\rm MA}(Q^2) 
= \frac{1}{\pi} \int_{0}^{+\infty} \, 
\frac{ d s }{(s + t)} \, r^{(1)}_{\rm pt}(s),~~ r^{(1)}_{\rm pt}(s)= {\rm Im} \; a_s^{(1)}(-s - i \epsilon) \,.
\label{disp_MA_LO}
\ee

This approach is commonly referred to as the {\it Minimal Approach} (MA) (see, e.g., \cite{Cvetic:2008bn}) or {\it Analytical Perturbation Theory} (APT) \cite{ShS}.

A further development of APT is the fractional APT (FAPT) \cite{BMS1}, which extends the construction principles to PT series involving non-integer powers of the coupling. In quantum field theory, such series arise for quantities with non-zero anomalous dimensions.

In this brief paper, we summarize the fundamental properties of MA couplings, as derived in \cite{Kotikov:2022sos} using the $1/L$-expansion. For the standard coupling, this expansion is only valid for large $Q^2$, i.e., $Q^2 \gg \Lambda^2$. However, as demonstrated in \cite{Kotikov:2022sos,KoZe23}, the situation is entirely different for analytic couplings: the $1/L$-expansion is applicable for all values of the argument. This is because non-leading corrections in the expansion vanish not only as $Q^2 \to \infty$ but also as $Q^2 \to 0$\footnote{This observation was previously made in \cite{ShS}.}, resulting only in small, non-zero corrections in the region $Q^2 \sim \Lambda^2$.

Below, we first present the LO results, followed by the NLO results, building on our previous results (\ref{tdmp1N}) for the standard strong coupling.

{\bf LO.}~~
The LO MA coupling $A^{(1)}_{{\rm MA},\nu,0}$ takes the form \cite{BMS1}:
\be
A^{(1)}_{{\rm MA},\nu,0}(Q^2) = {\left( a^{(1)}_{\nu,0}(Q^2)\right)}^{\nu} - \frac{{\rm Li}_{1-\nu}(z_0)}{\Gamma(\nu)}
\equiv \frac{1}{L_0^{\nu}}-\Delta^{(1)}_{\nu,0}\,,
\label{tAMAnu}
\ee
where
\be
   {\rm Li}_{\nu}(z)=\sum_{m=1}^{\infty} \, \frac{z^m}{m^{\nu}}=  \frac{z}{\Gamma(\nu)} \int_0^{\infty} 
\frac{ dt \; t^{\nu -1} }{(e^t - z)}
   \label{Linu}
\ee
is the Polylogarithm.

For $\nu=1$, we recover the well-known result of Shirkov and Solovtsov \cite{ShS}:
\be
A^{(1)}_{\rm MA,0}(Q^2) \equiv A^{(1)}_{\rm MA,\nu=1,0}(Q^2)
=\frac{1}{L_0}- \frac{z_0}{1-z_0}\,,
\label{tAM1}
\ee
This result can be directly obtained from the integral forms (\ref{disp_MA_LO}).

{\bf NLO.}~~ By analogy with the standard coupling and using the results (\ref{tdmp1N}), we obtain for the MA analytic coupling $\tilde{A}^{(i+1)}_{{\rm MA},\nu,i}$ the expressions:
\be
\tilde{A}^{(2)}_{{\rm MA},\nu,1}(Q^2) = \tilde{A}^{(1)}_{{\rm MA},\nu,1}(Q^2) +
\nu \, \tilde{\delta}^{(2)}_{{\rm MA},\nu,i}(Q^2),
\label{tAiman}
\ee
where $\tilde{A}^{(1)}_{{\rm MA},\nu,i}$ is given in Eq. (\ref{tAMAnu}) and
\be
\tilde{\delta}^{(2)}_{{\rm MA},\nu,1}(Q^2)= \tilde{\delta}^{(2)}_{\nu,1}(Q^2) -\hat{R}_1
\left( \frac{{\rm Li}_{-\nu}(z_1)}{\Gamma(\nu+1)}\right)=\tilde{\delta}^{(2)}_{\nu,1}(Q^2) - \Delta^{(2)}_{\nu,1}(z_1),
\label{tdAman}
\ee
with $\overline{\gamma}_{\rm E}=\gamma_{\rm E}-1$,
\be
\Delta^{(2)}_{\nu,1}(z)
=b_1\Bigl[\overline{\gamma}_{\rm E}
  {\rm Li}_{-\nu}(z)+{\rm Li}_{-\nu,1}(z)\Bigr],~~
 {\rm Li}_{\nu,1}(z)= \sum_{m=1}  \,\frac{z^m \,\ln m}{m^{\nu}},~~
 {\rm Li}_{-1}(z)= \frac{z}{(1-z)^2}
    \, .
\label{Lii.1}
\ee
and $\tilde{\delta}^{(2)}_{\nu,1}(Q^2)$ and $\hat{R}_1$ are given in Eqs. (\ref{tdmp1N}) and (\ref{L}), respectively.

\begin{figure}[t]
\centering
\includegraphics[width=0.58\textwidth]{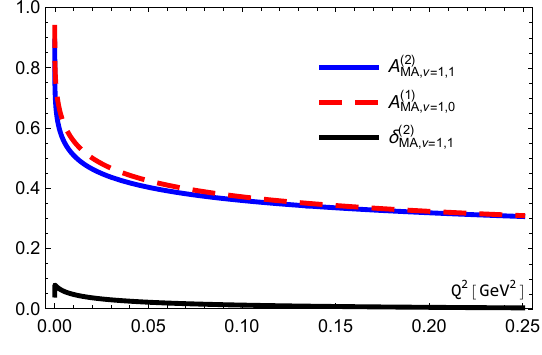}
\caption{\label{fig:Adelta2}
      Results for $A^{(1)}_{\rm MA,\nu=1,0}(Q^2)$, $A^{(2)}_{\rm MA,\nu=1,1}(Q^2)$, and $\delta^{(2)}_{\rm MA,\nu=1,1}(Q^2)$.
    }
\end{figure}

The analytical results for the MA coupling $\tilde{A}^{(i+1)}_{{\rm MA},\nu,i}$ can be explicitly found for $\nu=1$.

Figure \ref{fig:Adelta2} shows that $A^{(i+1)}_{\rm MA,i}(Q^2)$ are very close to each other for $i=0$ and $i=1$. The differences $\delta^{(2)}_{\rm MA,\nu=1,1}(Q^2)$ between the LO and NLO results are non-zero only for $Q^2 \sim \Lambda^2$.

\section{MA Coupling Form Convenient at $Q^2 \sim \Lambda^2$}

The results (\ref{tAMAnu}) and (\ref{tAiman}) for MA couplings are very convenient in the regimes of large and small $Q^2$. However, for $Q^2 \sim \Lambda_i^2$, both the standard coupling and the additional term $\delta^{(i+1)}_{\rm MA,\nu,i}(Q^2)$ contain singularities that cancel in the sum. Consequently, numerical applications of these results may be challenging, potentially requiring sub-expansions for each part near $Q^2 =\Lambda_i^2$. Therefore, we propose an alternative form that is particularly useful for $Q^2 \sim \Lambda_i^2$ and can also be used for other $Q^2$ values, except in the extremes of very large or very small $Q^2$.

{\bf LO.}~~
The LO MA coupling $A^{(1)}_{{\rm MA},\nu}(Q^2)$ \cite{ShS} can also be expressed as \cite{BMS1}:
\be
A^{(1)}_{{\rm MA},\nu}(Q^2) =
\frac{(-1)}{\Gamma(\nu)} \, \sum_{r=0}^{\infty} \zeta(1-\nu-r) \, \frac{(-L)^r}{r!}~ (L<2\pi),~~\zeta(\nu)
     =\sum_{m=1}^{\infty} \, \frac{1}{m^{\nu}}
\label{tAMAnuNew}
\ee
where $\zeta(\nu)$ denotes the Euler $\zeta$-function.

The result (\ref{tAMAnuNew}) was derived in Ref. \cite{BMS1} using properties of the Lerch function, which generalizes the Polylogarithms (\ref{Linu}).

For $\nu=1$, we have:
\be
A^{(1)}_{\rm MA}(L) = - \, \sum_{r=0}^{\infty} \zeta(-r) \, \frac{(-L)^r}{r!},~~
\zeta(-2m) = -\frac{\delta^0_m}{2},~~\zeta(-(1+2l))= - \frac{B_{2(l+1)}}{2(l+1)},
\label{tAMA1New}
\ee
where $\delta^0_m$ is the Kronecker delta and $B_{r+1}$ are Bernoulli numbers.

{\bf NLO.}~~
Now consider the derivatives of the MA coupling, $\tilde{A}^{(1)}_{{\rm MA},\nu}$, given in Eq. (\ref{tAiman}):
\be
\tilde{A}^{(2)}_{{\rm MA},\nu,1}(Q^2) = \tilde{A}^{(1)}_{{\rm MA},\nu,1}(Q^2) +
\nu\,\tilde{\delta}^{(2)}_{{\rm MA},\nu,1}(Q^2)\,,~~
\tilde{\delta}^{(2)}_{{\rm MA},\nu,1}(Q^2)= \hat{R}_1   \, A^{(1)}_{{\rm MA},\nu+1,1} \, ,
\label{tdAmanNew}
\ee
where the operator $\hat{R}_1$ is defined in (\ref{hR_i}).

After some calculations, we obtain:
\be
\tilde{\delta}^{(2)}_{{\rm MA},\nu,1}(Q^2)=
\frac{(-1)}{\Gamma(\nu+1)} \, \sum_{r=0}^{\infty} \tilde{R}_1(\nu+r) \, \frac{(-L_1)^r}{r!}
\label{tdAmanNew}
\ee
where
\be
\tilde{R}_1(\nu+r)=b_1\Bigl[\overline{\gamma}_{\rm E}
  \zeta(-\nu-r)+\zeta_1(-\nu-r)\Bigr],~~
\zeta_k(\nu)=
\sum_{m=1}^{\infty} \, \frac{\ln^k m}{m^{\nu}}
\, .
   \label{zetaknu}
\ee

The functions $\zeta_n(-\nu-r)$ are not well-defined for large $r$, so we replace them using:
\be
\zeta(-\nu-r)= -\frac{\Gamma(\nu+r+1)}{\pi(2\pi)^{\nu+r}}\,\tilde{\zeta}(\nu+r+1),~~\tilde{\zeta}(\nu+r+1)=\sin\left[\frac{\pi}{2}(\nu+r)\right] \, \zeta(\nu+r+1) \, .
  \label{ze_nu.1N}
\ee
After further calculations, we find:
\be
\tilde{\delta}^{(2)}_{{\rm MA},\nu,k}(Q^2)=
\frac{1}{\Gamma(\nu+1)} \, \sum_{r=0}^{\infty} \frac{\Gamma(\nu+r+1)}{\pi(2\pi)^{\nu+r}} \, Q_1(\nu+r+1) 
\, \frac{(-L_k)^r}{r!}\,,
\label{tdAmanNew}
\ee
where
\be
Q_1(\nu+r+1)=b_1\Bigl[\tilde{Z}_1(\nu+r)\tilde{\zeta}(\nu+r+1)+\tilde{\zeta}_1(\nu+r+1)\Bigr],~~
\tilde{Z}_1(\nu)=\hat{Z}_1(\nu)-\ln(2\pi)\,,~~
\label{Qi}
\ee

Using the definition of $\tilde{\zeta}(\nu)$ from (\ref{ze_nu.1N}), we have:
\be
\tilde{\zeta}_1(\nu+r+1)=\sin\left[\frac{\pi}{2}(\nu+r)\right] \, \zeta_1(\nu+r+1) + \frac{\pi}{2} \,
\cos\left[\frac{\pi}{2}(\nu+r)\right] \, \zeta(\nu+r+1) \, ,
\label{tzeta1234}
\ee
where $\zeta_1(\nu)$ is given in Eq. (\ref{zetaknu}).

Thus, we can rewrite the results (\ref{tdAmanNew}) as:
\be
Q_{1}(\nu+r+1)=\sin\left[\frac{\pi}{2}(\nu+r)\right] Q_{1a}(\nu+r+1)
+ \frac{\pi}{2} \,
\cos\left[\frac{\pi}{2}(\nu+r)\right] Q_{1b}(\nu+r+1)\, ,
\label{Qmab}
\ee
where
\be
Q_{1a}(\nu+r+1)=b_1\Bigl[\tilde{Z}_1(\nu+r)\zeta(\nu+r+1)+\zeta_1(1,\nu+r+1)\Bigr],~~
Q_{1b}(\nu+r+1)=b_1\zeta(\nu+r+1),
\label{Qiab}
\ee

The results for the MA coupling itself are obtained by setting $\nu=1$. Moreover, at $L_k=0$, i.e., for $Q^2=\Lambda_k^2$, we find:
\be
A^{(1)}_{\rm MA}= \frac{1}{2},~~~
\delta^{(2)}_{s}=
-\frac{b_1}{2\pi^2} \, \Bigl(\zeta_1(2)+l\zeta(2)\ln(2\pi)\Bigr),~~
\label{dAmanNew2}
\ee

\section{Conclusions}

In this short paper, we have summarized the results from our recent work \cite{Kotikov:2022sos}. In particular, \cite{Kotikov:2022sos} provides $1/L$-expansions for the $\nu$-derivatives of the strong coupling $a_s$, expressed as combinations of the operators $\hat{R}_i$ (\ref{hR_i}) applied to the LO coupling $a_s^{(1)}$. By applying the same operators to the $\nu$-derivatives of the LO MA coupling $A_{\rm MA}^{(1)}$, four different representations for $\tilde{A}_{\rm MA,\nu}^{(i)}$ were obtained at each $i$-th order of PT. All results are presented in \cite{Kotikov:2022sos,KoZe23} up to the 5th order of PT, where the corresponding QCD $\beta$-function coefficients are well known (see \cite{Baikov:2016tgj}). In this paper, we have restricted ourselves to the first two orders to avoid the more cumbersome results from the higher orders.

For the MA coupling, higher-order corrections are negligible in both the $Q^2 \to 0$ and $Q^2 \to \infty$ limits, and are only non-zero in the vicinity of $Q^2 =\Lambda^2$. Thus, they represent only minor corrections to the LO MA coupling $A_{\rm MA}^{(1)}(Q^2)$.

The results of \cite{Kotikov:2022sos} have recently been successfully applied to studies of the (polarized) Bjorken sum rule \cite{Gabdrakhmanov:2023rjt} and the Gross-Llewellyn Smith sum rule \cite{GLS} (see also the review in \cite{Gabdrakhmanov:2025afi}). For these studies, the form of the MA coupling convenient for $Q^2\sim\Lambda^2$ was extensively used (as reviewed in Section 5).

{\bf Acknowledgments}~
We thank Hannes Jung for the invitation to participate in the workshop "QCD at the Extremes".
One of us (I.A.Z.)
is supported by Fellowship ANID Beca de Doctorado Nacional No. 21250067.

\end{document}